\documentstyle[12pt,indent,eqsection,subeqnarray]{article}

\begin{document}

\bibliographystyle{plain}

\title{ The 3-State Square-Lattice Potts Antiferromagnet at
        Zero Temperature
      }
\author{
  {\small Jes\'us Salas}                  \\[-0.2cm]
  {\small\it Departamento de F\'{\i}sica de la Materia Condensada and} \\[-0.2cm]
  {\small\it Departamento de F\'{\i}sica Te\'orica} \\[-0.2cm]
  {\small\it Facultad de Ciencias}        \\[-0.2cm]
  {\small\it Universidad de Zaragoza}     \\[-0.2cm]
  {\small\it 50009 Zaragoza SPAIN}       \\[-0.2cm]
  {\small\tt JESUS@JUPITER.UNIZAR.ES}     \\[5mm]
  {\small Alan D.~Sokal}                  \\[-0.2cm]
  {\small\it Department of Physics}       \\[-0.2cm]
  {\small\it New York University}         \\[-0.2cm]
  {\small\it 4 Washington Place}          \\[-0.2cm]
  {\small\it New York, NY 10003 USA}      \\[-0.2cm]
  {\small\tt SOKAL@NYU.EDU}               \\[-0.2cm]
  {\protect\makebox[5in]{\quad}}  
  \\
}
\vspace{0.5cm}

\maketitle
\thispagestyle{empty}   

\def\spose#1{\hbox to 0pt{#1\hss}}
\def\ltapprox{\mathrel{\spose{\lower 3pt\hbox{$\mathchar"218$}}
 \raise 2.0pt\hbox{$\mathchar"13C$}}}
\def\gtapprox{\mathrel{\spose{\lower 3pt\hbox{$\mathchar"218$}}
 \raise 2.0pt\hbox{$\mathchar"13E$}}}
\def\inapprox{\mathrel{\spose{\lower 3pt\hbox{$\mathchar"218$}}
 \raise 2.0pt\hbox{$\mathchar"232$}}}

\vspace{1cm}

\begin{abstract}
We study the 3-state square-lattice Potts antiferromagnet
at zero temperature by a Monte Carlo simulation
using the Wang-Swendsen-Koteck\'y cluster algorithm,
on lattices up to $1024 \times 1024$.
We confirm the critical exponents predicted
by Burton and Henley based on the height representation of this model.
\end{abstract}


\clearpage

\newcommand{\be}{\begin{equation}}
\newcommand{\ee}{\end{equation}}
\newcommand{\<}{\langle}
\renewcommand{\>}{\rangle}
\newcommand{\para}{\|}
\renewcommand{\perp}{\bot}

\def\smfrac#1#2{{\textstyle\frac{#1}{#2}}}
\def\half{ {{1 \over 2 }}}
\def\smhalf{ {\smfrac{1}{2}} }
\def\scra{{\cal A}}
\def\scrc{{\cal C}}
\def\scrd{{\cal D}}
\def\scre{{\cal E}}
\def\scrf{{\cal F}}
\def\scrh{{\cal H}}
\def\scri{{\cal I}}
\def\scrk{{\cal K}}
\def\scrm{{\cal M}}
\newcommand{\scrmvec}{\vec{\cal M}_V}
\def\scrmtens{{\stackrel{\leftrightarrow}{\cal M}_T}}
\def\scro{{\cal O}}
\def\scrp{{\cal P}}
\def\scrr{{\cal R}}
\def\scrs{{\cal S}}
\def\ttens{{\stackrel{\leftrightarrow}{T}}}
\def\scrv{{\cal V}}
\def\scrw{{\cal W}}
\def\scry{{\cal Y}}
\def\scrz{{\cal Z}}
\def\tauss{\tau_{int,\,\scrm^2}}
\def\taux{\tau_{int,\,{\cal M}^2}}
\newcommand{\taum}{\tau_{int,\,\vec{\cal M}}}
\def\taue{\tau_{int,\,{\cal E}}}
\newcommand{\imag}{\mathop{\rm Im}\nolimits}
\newcommand{\real}{\mathop{\rm Re}\nolimits}
\newcommand{\tr}{\mathop{\rm tr}\nolimits}
\newcommand{\sgn}{\mathop{\rm sgn}\nolimits}
\newcommand{\codim}{\mathop{\rm codim}\nolimits}
\newcommand{\rank}{\mathop{\rm rank}\nolimits}
\newcommand{\sech}{\mathop{\rm sech}\nolimits}
\def\textprime{{${}^\prime$}}
\newcommand{\longto}{\longrightarrow}
\def\var{ \hbox{var} }
\newcommand{\gtilde}{ {\widetilde{G}} }
\newcommand{\USp}{ \hbox{\it USp} }
\newcommand{\CP}{ \hbox{\it CP\/} }
\newcommand{\QP}{ \hbox{\it QP\/} }
\def\hboxscript#1{ {\hbox{\scriptsize\rm #1}} }

\newcommand{\plotdot}{\makebox(0,0){$\bullet$}}
\newcommand{\plotsmalldot}{\makebox(0,0){{\footnotesize $\bullet$}}}

\def\bsigma{\mbox{\protect\boldmath $\sigma$}}
\def\bpi{\mbox{\protect\boldmath $\pi$}}
\def\btau{\mbox{\protect\boldmath $\tau$}}
\def\bn{{\bf n}}
\def\br{{\bf r}}
\def\bz{{\bf z}}
\def\bh{\mbox{\protect\boldmath $h$}}

\def\betatilde{ {\widetilde{\beta}} }
\def\hatp{\hat p}
\def\hatl{\hat l}

\def\msbar{ {\overline{\hbox{\scriptsize MS}}} }
\def\normalmsbar{ {\overline{\hbox{\normalsize MS}}} }

\def\eff{ {\hbox{\scriptsize\em eff}} }

\newcommand{\reff}[1]{(\ref{#1})}

\newcommand{\Z}{{\bf Z}}
\newcommand{\zed}{{\bf Z}}
\newcommand{\R}{{\bf R}}
\font\srm=cmr7 		
\def\szed{\hbox{\srm Z\kern-.45em\hbox{\srm Z}}}
\def\sR{\hbox{{\srm I}\kern-.2em\hbox{\srm R}}}
\def\C{{\bf C}}



\newtheorem{theorem}{Theorem}[section]
\newtheorem{corollary}[theorem]{Corollary}
\newtheorem{lemma}[theorem]{Lemma}
\newtheorem{conjecture}[theorem]{Conjecture}
\newtheorem{definition}[theorem]{Definition}
\def\proof{\bigskip\par\noindent{\sc Proof.\ }}
\def\qed{\hbox{\hskip 6pt\vrule width6pt height7pt depth1pt \hskip1pt}\bigskip}

%
%
\newenvironment{sarray}{
          \textfont0=\scriptfont0
          \scriptfont0=\scriptscriptfont0
          \textfont1=\scriptfont1
          \scriptfont1=\scriptscriptfont1
          \textfont2=\scriptfont2
          \scriptfont2=\scriptscriptfont2
          \textfont3=\scriptfont3
          \scriptfont3=\scriptscriptfont3
        \renewcommand{\arraystretch}{0.7}
        \begin{array}{l}}{\end{array}}

\newenvironment{scarray}{
          \textfont0=\scriptfont0
          \scriptfont0=\scriptscriptfont0
          \textfont1=\scriptfont1
          \scriptfont1=\scriptscriptfont1
          \textfont2=\scriptfont2
          \scriptfont2=\scriptscriptfont2
          \textfont3=\scriptfont3
          \scriptfont3=\scriptscriptfont3
        \renewcommand{\arraystretch}{0.7}
        \begin{array}{c}}{\end{array}}


\section{Introduction}  \label{sec_intro}

Antiferromagnetic Potts models \cite{Potts_52,Wu_82,Wu_84}
are much less well understood than their ferromagnetic counterparts.
One reason for this is that the behavior depends strongly
on the microscopic lattice structure,
in contrast to the universality typically enjoyed by ferromagnets.
As a result, many basic questions have to be investigated case-by-case:
Is there a phase transition at finite temperature, and if so, of what order?
What is the nature of the low-temperature phase?
If there is a critical point, what are the critical exponents and the
universality classes?
Can these exponents be understood (for two-dimensional models)
in terms of conformal field theory?

One thing is known rigorously \cite{Kotecky_88,Salas_Sokal_97}:
for $q$ large enough (how large depends on the lattice in question),
the antiferromagnetic $q$-state Potts model has
a unique infinite-volume Gibbs measure and
exponential decay of correlations
at all temperatures, {\em including zero temperature}\/:
the system is disordered as a result of the
large ground-state entropy.\footnote{
   This behavior has been proven for
   $q\geq4$ on the hexagonal lattice,
   $q\geq6$ on the Kagom\'e lattice,
   $q\geq7$ on the square lattice, and
   $q\geq11$ on the triangular lattice \cite{Salas_Sokal_97}.
   However, these bounds are presumably {\em not}\/ sharp: see
   equation \protect\reff{qc} below.
}
However, for smaller values of $q$, phase transitions can and do occur.
One expects that for each lattice ${\cal L}$ there
will be a value $q_c({\cal L})$ such that
\begin{itemize}
   \item[(a)]  For $q > q_c({\cal L})$  the model has exponential decay
       of correlations uniformly at all temperatures,
       including zero temperature.
   \item[(b)]  For $q = q_c({\cal L})$  the model has a critical point
       at zero temperature.
   \item[(c)]  For $q < q_c({\cal L})$  any behavior is possible.
       Often (though not always) the model has a phase transition
       at nonzero temperature, which may be of either first or second order.
\end{itemize}
The problem, for each lattice, is to find $q_c({\cal L})$
and to determine the precise behavior for each $q \le q_c({\cal L})$.

For the common two-dimensional lattices,
strong theoretical arguments\footnote{
   Summarized in the introduction to \cite{Salas_Sokal_97}.
}
--- which, however, fall short of a rigorous proof ---
yield the following predictions for $q_c({\cal L})$:
\be
   q_c({\cal L})   \;=\;
   \cases{ (3+\sqrt{5})/2 \approx 2.618\ldots  & for the hexagonal lattice \cr
           3                                   & for the square lattice \cr
           3                                   & for the Kagom\'e lattice \cr
           4                                   & for the triangular lattice \cr
         }
 \label{qc}
\ee
Monte Carlo simulations have confirmed
numerically that the 3-state square-lattice model has a zero-temperature
critical point \cite{Ferreira_Sokal_95,Ferreira_Sokal_prep},
and that the 4-state square-lattice model
\cite{Ferreira_Sokal_95,Ferreira_Sokal_prep}
and the 3-state hexagonal-lattice model \cite{Shrock_95,Salas_prep} are
non-critical uniformly down to zero temperature.\footnote{
   The Monte Carlo simulations of the 3-state hexagonal-lattice model
   reported in Ref.~\cite{Shrock_95} give no evidence of any
   {\em first-order}\/ phase transition as the temperature is
   varied from infinity to zero;
   this behavior is consistent with the theoretical prediction
   that the model has exponential decay of correlations
   uniformly down to zero temperature.
   However, these authors did not measure the correlation length
   or the staggered susceptibility, so no direct test of the
   non-criticality at zero temperature was made.
   Such a direct test is being made in Ref.~\cite{Salas_prep}.
}

%
%
%

Two-dimensional models with zero-temperature critical points
are of particular interest,
as they can in most cases be mapped onto a ``height''
(or ``interface'' or ``SOS-type'') model
\cite{Nijs_82,Blote_82,Nienhuis_84,Kolafa_84,Levitov_90,Thurston_90,%
Huse_92,Blote_94,Kondev_94,Kondev_95,Kondev_96,Kondev_96b,%
Kondev_96c,Raghavan_97,Zeng_97,Henley_97,Burton_Henley_97,Henley_unpublished}.
If this height model lies in its ``rough'' phase
--- a question that has to be investigated on a case-by-case basis ---
then its long-distance behavior is that of a massless Gaussian
with some ({\em a priori}\/ unknown) ``stiffness'' $K > 0$.
The critical operators can then be identified via the height mapping,
and the corresponding critical exponents can be predicted in terms
of the single parameter $K$.
In particular, if we know (by some other means) one of these exponents,
then we can deduce the rest.

Height representations thus give a means for recovering
a sort of universality for some (but not all)
antiferromagnetic Potts models
and for understanding their critical behavior
in terms of conformal field theory.
All the nonuniversal details of the microscopic lattice structure
are encoded in the height representation
and in the stiffness parameter $K$.
Given these, everything can be understood in terms of
the universal behavior of massless Gaussian fields.

The plan of this paper is as follows:
In Section~\ref{sec_height_representations} we present briefly
the general theory of height representations
and then work out in detail the case of the
3-state square-lattice Potts antiferromagnet.
Our presentation is based on the work of Henley and collaborators
\cite{Kondev_95,Kondev_96,Raghavan_97,Zeng_97,Burton_Henley_97,%
Henley_unpublished},
supplemented by a few minor innovations of our own.
In the remainder of the paper, our goal is to test,
by Monte Carlo simulation,
the critical exponents predicted
by Burton and Henley \cite{Burton_Henley_97}
for the three relevant operators
in the 3-state square-lattice Potts antiferromagnet at zero temperature.
In Section~\ref{sec_numerical_simulations}
we describe our simulations,
and in Section~\ref{sec_data_analysis} we analyze the data.

\section{Height Representations}   \label{sec_height_representations}

Many two-dimensional models with zero-temperature critical points
can be mapped onto a ``height'' model:
these include
the triangular-lattice Ising antiferromagnet \cite{Blote_82,Nienhuis_84},
the triangular-lattice spin-$S$ Ising antiferromagnet \cite{Zeng_97},
the 3-state square-lattice Potts antiferromagnet
\cite{Nijs_82,Kolafa_84,Burton_Henley_97},
the 3-state Kagom\'e-lattice Potts antiferromagnet
\cite{Huse_92,Kondev_96},
the 4-state triangular-lattice Potts antiferromagnet
\cite{Henley_unpublished},
the 4-state Potts antiferromagnet on the covering lattice of the square lattice
\cite{Kondev_95,Kondev_96},
a constrained 4-state Potts antiferromagnet on the square lattice
\cite{Burton_Henley_97},
a special 6-vertex model \cite{Kondev_96},
and various dimer models
\cite{Levitov_90,Thurston_90,Raghavan_97,Henley_97}
and fully packed loop models
\cite{Blote_94,Kondev_94,Kondev_96b,Kondev_96c}.
Here we shall explain briefly the basic principles
underlying the construction of such mappings
and their use to extract critical exponents.
We shall then work out in detail the case of the
3-state square-lattice Potts antiferromagnet.

\subsection{General Theory}

The first step
is to define a map assigning to each
zero-temperature spin configuration $\{\sigma(x)\}$ a corresponding
microscopic height configuration $\{h(x)\}$.
This {\em height rule}\/ is usually defined by local increments,
i.e.\ one prescribes the change $\Delta h \equiv h(y)- h(x)$
in going from a site $x$ to a neighboring site $y$
in terms of the spin variables $\sigma(x)$ and $\sigma(y)$.
For such a rule to be well-defined, one must verify that in all cases
the net increment $\Delta h$ around any closed loop is zero.\footnote{
   In fact, this property usually holds for free boundary conditions
   but {\em not}\/ for periodic boundary conditions.
   We shall hereafter neglect this latter subtlety,
   by imagining that we are working always in infinite volume.
}
The height variables $h(x)$ lie in some discrete set
$\scrh \subset {\bf R}^D$ (for some suitable dimension $D$),
which we call the {\em height lattice}\/.

The next step is to identify the so-called {\em ideal states}\/:
these are ground-state configurations (or families of configurations)
of the original spin model
whose corresponding height configurations are macroscopically ``flat''
(i.e.\ have zero net slope)
and which maximize the entropy density
(in the sense of maximizing the number of ground states that can be
obtained from the ideal states by local modifications of the spins).
We label each ideal state by its average height $h \in {\bf R}^D$,
and we define the
{\em ideal-state lattice}\/ ${\cal I} \subset {\bf R}^D$
to be the set of all average heights of ideal states.
The {\em equivalence lattice}\/
\be
   {\cal E}  \;=\;  \{ a \in {\bf R}^D \colon\; a + {\cal I} = {\cal I} \}
\ee
is the subgroup of ${\bf R}^D$ summarizing the underlying periodicity
of ${\cal I}$.

We now guess that, in typical configurations of the spin model,
the lattice is subdivided into reasonably large domains
in which the spin configuration closely resembles one of the ideal states.
It follows that typical configurations of the height model
are given by domains in which the height $h(x)$ exhibits small fluctuations
around one of the values in the ideal-state lattice.
We therefore expect that a suitably defined coarse-grained height variable
$\bar{h}(x)$ will take values in or near the ideal-state lattice ${\cal I}$,
except at boundaries between domains.
The long-wavelength behavior of the height model is thus postulated to be
controlled by an effective coarse-grained Hamiltonian of the form
\begin{equation}
H \;=\; \int d^2x \, \left[ {K \over 2} \sum_{i=1}^D |\nabla \bar{h}_i|^2
                     + V_{\rm lock}(\bar{h}(x)) \right]
  \;,
 \label{macro_H}
\end{equation}
where we have made explicit the components of the macroscopic height
$\bar{h} = (\bar{h}_1,\bar{h}_2,\ldots,\bar{h}_D)$.
The gradient term in \reff{macro_H} takes into account
the entropy of small fluctuations around the ideal states;
the second term is the so-called {\em locking potential}\/,
which favors the heights to take their values in ${\cal I}$.
We then expect that there exists some constant $K_r$ such that
for $K < K_r$ (resp.\ $K > K_r$) the locking potential is
irrelevant (resp.\ relevant) in the renormalization-group sense.
Thus, if $K < K_r$ our surface model is ``rough'' and its
long-wavelength behavior can be described by a massless
Gaussian model \cite{Ginsparg,Kondev_96} with $D$ components:
\be
   \Bigl\langle [ \bar{h}_i(x) - \bar{h}_i(y)] \,
                [ \bar{h}_j(x) - \bar{h}_j(y)]  \Bigr\rangle
   \;\approx\;  {\delta_{ij} \over \pi K} \, \log |x-y|
\ee
for $|x-y|\gg 1$;
in this case, the original zero-temperature spin system is critical.
If, on the other hand, $K>K_r$, then the surface model is in its
``smooth'' phase, exhibiting long-range order
\be
\langle  \bar{h}(x) \rangle  \;=\;  {\bf h}_0
\ee
and {\em bounded}\/ fluctuations around this ordered state:
\be
   \Bigl\langle [\bar{h}(x) - \bar{h}(y)]^2   \Bigr\rangle
   \; \hbox{ bounded as } |x-y| \to \infty  \;;
\ee
correspondingly, the spin system is ``locked''
into small fluctuations around one of the ideal states.
At $K=K_r$ the surface model undergoes a roughening
transition.\footnote{
   Note that the only alternatives for the spin model
   are criticality and long-range order.
   Thus, if there exists a height representation,
   the original spin model cannot be disordered at zero temperature.
}

Let us note, finally, that a given ideal state can be represented by
many different average heights $h \in {\cal I}$.
More precisely, suppose that in some domain we have a particular ideal state X
and that its average height is $h_0 \in {\cal I}$.
Now let us pass through various other domains of the lattice,
coming back finally to a domain in which the ideal state is again X.
We will find the average height in this latter domain to lie in the set
$h_0 + {\cal R}$, where ${\cal R}$ is a particular subgroup of ${\cal E}$
that we call the {\em repeat lattice}\/;
we will also find, conversely, that whenever we enter a domain in which
the average height lies in $h_0 + {\cal R}$,
that domain is in ideal state X.
It follows that the ideal states are in one-to-one correspondence
with the cosets ${\cal I}/{\cal R}$.

The coarse-grained correlation functions of
local operators in the spin language
can be understood in terms of the correlation functions of
local operators of the coarse-grained heights.
The important point is that these latter operators
should have the periodicity of the repeat lattice ${\cal R}$.
This means that the Fourier expansion of such an operator ${\bf O}$,
\begin{equation}
   {\bf O}(x) \;=\;
   \sum_{G \in {\cal R}^\circ} O_G \, e^{i G \cdot \bar{h}(x)}
   \;,
 \label{fourier}
\end{equation}
contains only wavevectors belonging to
the reciprocal lattice of the repeat lattice,
\be
   {\cal R}^\circ   \;\equiv\;
   \{ G \in {\bf R}^D \colon\;  G \cdot a \in 2\pi \zed
      \hbox{ for all } a \in {\cal R}   \}
   \;.
\ee
On the other hand, two wavevectors whose difference belongs to
the reciprocal lattice of the equivalence lattice,
\be
   {\cal E}^\circ   \;\equiv\;
   \{ G \in {\bf R}^D \colon\;  G \cdot a \in 2\pi \zed
      \hbox{ for all } a \in {\cal E}   \}
   \;,
\ee
give rise to vertex operators $\exp[i G \cdot \bar{h}(x)]$
having identical long-distance behavior.
The vertex operators of the height model
are thus in one-to-one correspondence with the cosets
${\cal R}^\circ/{\cal E}^\circ  \simeq  ({\cal E}/{\cal R})^\circ$.

Now, provided that the height model is in the rough phase ($K<K_r$),
the correlation functions of the vertex operators
$\exp[i G \cdot \bar{h}(x)]$ are given by
\be
  \left\langle  e^{iG \cdot \bar{h}(0)} \, e^{-iG' \cdot \bar{h}(x)}
  \right\rangle
  \;=\;  \delta_{G,G'} \, \exp\left[ -{G^2 \over 2}
     \Bigl\langle [\bar{h}(0)-\bar{h}(x)]^2 \Bigr\rangle  \right]
  \;\sim\;  \delta_{G,G'} \, |x|^{-G^2/(2\pi K)}
  \;.
\ee
It follows that the critical behavior of the operator ${\bf O}$ will
be given by the most relevant vertex operator $\exp[i G \cdot \bar{h}(x)]$
appearing in its Fourier expansion with a nonzero coefficient:
\begin{equation}
  \langle {\bf O}(0) \, {\bf O}(x)^* \rangle
  \;\sim\;  |x|^{-\eta_{\bf O}}
\end{equation}
where
\begin{equation}
  \eta_{\bf O} \;=\;  \min\limits_{O_G \neq 0}  {G^2 \over 2 \pi K}  \;\,.
\end{equation}
(The scaling dimension is $x_{\bf O} = \eta_{\bf O}/2$,
 and the operator is relevant in case the
 renormalization-group eigenvalue $d - x_{\bf O} = 2 - x_{\bf O}$
 is $> 0$.)
This formula implies that we can write {\em all}\/ the
critical exponents in terms of a single parameter $K$.
If one exponent is known, then all of them are.

In particular, the locking potential $V_{\rm lock}$
has the periodicity of the ideal-state lattice ${\cal I}$;
its Fourier expansion \reff{fourier} has contributions
only from wavevectors $G$ belonging to ${\cal E}^\circ$.
Let $a_{{\cal E}^\circ}$ be the length of the smallest nonzero vector
in ${\cal E}^\circ$.
Now, the roughening transition occurs exactly where the locking potential
is marginal, i.e.\ where $\eta_{V_{\rm lock}} = 4$.
It follows that
\be
   K_r  \;=\;  {a_{{\cal E}^\circ}^2  \over  8\pi }    \;\,.
\ee
If $K<K_r$, the locking potential is irrelevant,
with scaling dimension
\be
   x_{V_{\rm lock}} \;=\;  2 K_r/K  \;>\;  2   \;.
\ee
It induces corrections to scaling $\sim L^{2 - x_{V_{\rm lock}}}$,
where $L$ is a suitable length scale.

In addition to vertex operators $\exp[i G \cdot \bar{h}(x)]$,
there is another type of local operator that makes sense in the
massless Gaussian model:  powers of gradients of $\bar{h}$.\footnote{
   The height $\bar{h}$ itself is ill-defined as a field
   in dimension $d \le 2$, due to infrared divergences.
   But gradients $\nabla \bar{h}$, $\nabla\nabla \bar{h}$, \ldots\
   are well-defined.
}
In particular, the operator $(\nabla \bar{h})^{2n}$
has scaling dimension 
\be
   x_{(\nabla \bar{h})^{2n}} \;=\;  2n
\ee
and hence $\eta_{(\nabla \bar{h})^{2n}} = 4n$.
It follows that all these operators are irrelevant, except the operator
$(\nabla \bar{h})^2$, which is marginal.
Since these operators respect the lattice symmetries,
they can appear in the effective Hamiltonian
and thereby induce corrections to scaling.
The leading such operator is $(\nabla \bar{h})^4$,
with scaling dimension $x_{(\nabla \bar{h})^4}= 4$;
it induces corrections
$\sim L^{2-x_{(\nabla \bar{h})^4}} = L^{-2}$.

Assuming that we have not overlooked any irrelevant operators
that could appear in the effective Hamiltonian,
we conclude that the leading corrections to scaling
behave as $L^{-\Delta}$, with
\be
   \Delta   \;=\;   \min\!\left( x_{V_{\rm lock}} - 2, \,
                                 x_{(\nabla \bar{h})^4} - 2  \right)
            \;=\;   \min\!\left( {2 K_r \over K} - 2 ,\, 2 \right)
   \;.
  \label{irrelevant}
\ee

We remark, finally, the height representation can also be applied to
these models at {\em nonzero}\/ temperature.
In that case one must consider also the fugacity of {\em defects}\/:
that is, of places where the zero-temperature constraints
are violated \cite{Kondev_96,Henley_unpublished}.
Very often the defect fugacity is a {\em relevant}\/ operator.

\subsection{Three-State Square-Lattice Potts Antiferromagnet}

The height representation of the 3-state square-lattice Potts antiferromagnet
at zero temperature is very simple
\cite{Nijs_82,Burton_Henley_97}.
Let the Potts spins $\sigma(x)$ take values in the set $\{0,1,2\}$.
The microscopic height variables $h(x)$ are then assigned as follows:
At the origin we take $h(0) = 0, 4, 2$ (mod~6) according as
$\sigma(0) = 0, 1, 2$;  this ensures that
\begin{subeqnarray}
     h(0) &=&   \sigma(0) \pmod{3}   \\
     h(0) &=&   0         \pmod{2}
  \label{h0}
\end{subeqnarray}
We then define the increment in height in going from a site $x$
to a nearest neighbor $y$ by
\begin{subeqnarray}
     h(x) - h(y) &=&   \sigma(x) - \sigma(y)   \pmod{3}   \\
     h(x) - h(y) &=&   \pm 1
  \label{hstep}
\end{subeqnarray}
This is well-defined (in free boundary conditions)
because the change $\Delta h$ around any plaquette is zero.\footnote{
   If four numbers $\pm 1$ add up to 0 mod~3, they must necessarily be
   two $+1$'s and two $-1$'s, hence add up to 0.
}
It follows from \reff{h0} and \reff{hstep} that
\begin{subeqnarray}
     h(x) &=&  \sigma(x)   \pmod{3}  \slabel{hmod3}  \\
     h(x) &=&  x_1 + x_2   \pmod{2}  \slabel{hmod2}
 \label{hx_mod6}
\end{subeqnarray}
for any site $x=(x_1,x_2)$.  In particular, the height $h(x)$
is uniquely determined mod~6 once we know the spin value $\sigma(x)$
and the parity of $x$, and conversely.
The height lattice $\scrh$ is clearly equal to $\zed$.

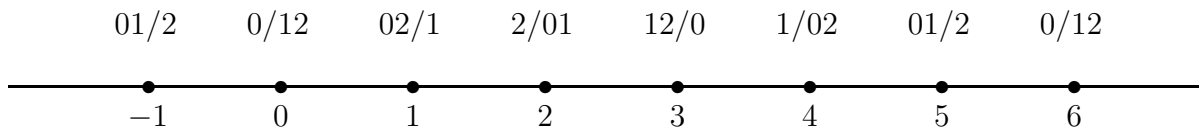
\begin{figure}
\centering
\begin{picture}(450,300)
   \thicklines
   \put(0,200){\line(1,0){450}}
   \put(50,197){$\bullet$}
   \put(100,197){$\bullet$}
   \put(150,197){$\bullet$}
   \put(200,197){$\bullet$}
   \put(250,197){$\bullet$}
   \put(300,197){$\bullet$}
   \put(350,197){$\bullet$}
   \put(400,197){$\bullet$}
   \put(45,185){$-1$}
   \put(100,185){0}
   \put(150,185){1}
   \put(200,185){2}
   \put(250,185){3}
   \put(300,185){4}
   \put(350,185){5}
   \put(400,185){6}
   \put(40,220){01/2}
   \put(90,220){0/12}
   \put(140,220){02/1}
   \put(190,220){2/01}
   \put(240,220){12/0}
   \put(290,220){1/02}
   \put(340,220){01/2}
   \put(390,220){0/12}
\end{picture}
\caption{
   Ideal-state lattice ${\cal I}$ for the 3-state square-lattice
   Potts antiferromagnet.
   The symbols above the graph indicate the ideal states
   of the spin model:
   0/12 means that the spins on the even sublattice are
   all equal to 0 and that the spins on the odd sublattice are chosen
   randomly between the values 1 and 2.
   The numbers below the graph indicate
   the average height for the given ideal state;
   this height is determined modulo 6.
}
\label{figure_graph}
\end{figure}

There are six ideal states, given by 0/12 (spins on the even sublattice
all equal to 0, spins on the odd sublattice chosen randomly
between 1 and 2) and its permutations.\footnote{
   States like 0/1 are not ideal states because
   they do not have maximal entropy density.
}
In an ideal state, the height is constant on the ordered sublattice
and fluctuates randomly $\pm 1$ around this level
on the disordered sublattice.
The average height of an ideal state is thus
equal to its height on the ordered sublattice;
it then follows from \reff{hx_mod6}
that there is a one-to-one correspondence between
ideal states and average heights mod~6
(see Figure~\ref{figure_graph}).\footnote{
   This fact was proven, in a different way, by Burton and Henley
   \cite[Appendix B.2]{Burton_Henley_97}.
}
The ideal-state lattice ${\cal I}$ is thus also $\zed$,
as is the equivalence lattice ${\cal E}$,
while the repeat lattice ${\cal R}$ is $6\zed$.
The corresponding reciprocal lattices are ${\cal E}^\circ = 2\pi\zed$
and ${\cal R}^\circ = (\pi/3) \zed$.

There are three relevant operators (in the renormalization-group sense)
appearing in this model (see Table~\ref{Table_operators})
\cite{Nijs_82,Burton_Henley_97}:

\begin{eqnarray}
{\bf M}_{stagg}(x)  &=&  (-1)^{x_1 + x_2} \, \vec{\sigma}(x)   \\[2mm]
{\bf M}_u(x)        &=&                      \vec{\sigma}(x)   \\[2mm]
{\bf P}_{stagg}(x)  &=&  {1 \over 4} (-1)^{x_1 + x_2}
   \sum_{y \hboxscript{ nnn of } x}  (2 \delta_{\sigma(x), \sigma(y)} - 1)
\end{eqnarray}
where we have represented the
Potts spin at site $x = (x_1,x_2)$ by a unit vector in the plane
\be
   \vec{\sigma}(x)  \;=\;  \left(\cos {2\pi \over 3}\sigma(x), \,
                                 \sin {2\pi \over 3} \sigma(x) \right)
   \;.
\ee
The first operator is the staggered magnetization;
the staggering corresponds to a momentum ${\bf k}_{stagg} =(\pi,\pi)$.
The second operator is the uniform magnetization.
The third operator is a staggered sum over
diagonal next-nearest-neighbor correlations
(i.e.\ over $y$ with $| y-x | = \sqrt{2}$);
we call it the staggered polarization.
In an ideal state, it takes the average value $+1$ (resp.\ $-1$)
according as it is the even (resp.\ odd) sublattice that is ordered.\footnote{
  Burton and Henley \cite{Burton_Henley_97} chose a slightly different
  definition of this operator:
  ${\bf P}_{stagg}(x) = (1/4)(-1)^{x_1 + x_2}
   \! \sum\limits_{y \hboxscript{ nnn of } x} \!
   \delta_{\sigma(x), \sigma(y)}$.
}

We can relate these observables directly to the {\em microscopic}\/
height variables $h(x)$ by exact identities.
For the vertex operators with $G = \pm \pi/3, \pm 2\pi/3, \pm \pi$ we have
\begin{eqnarray}
  e^{\pm i (\pi/3) h(x)}  & = &  M_{stagg}^{(1)}(x) \mp i M_{stagg}^{(2)}(x)
       \label{h_piover3}   \\[2mm]
  e^{\pm i (2\pi/3) h(x)} & = &  M_{u}^{(1)}(x) \pm i M_{u}^{(2)}(x)
       \label{h_2piover3}   \\[2mm]
  e^{\pm i \pi h(x)}      & = &  (-1)^{x_1 + x_2}
       \label{h_pi}
\end{eqnarray}
Here \reff{h_2piover3} and \reff{h_pi} follow immediately from
\reff{hmod3} and \reff{hmod2}, respectively,
while \reff{h_piover3} follows by multiplying these two
and taking the complex conjugate.
Of course, the strictly local operator \reff{h_pi} is trivial,
but we can define a nontrivial almost-local operator with
the same ($G=\pi$) long-distance behavior:
\be
   e^{\pm i (\pi/2) [h(x)+ h(y)]}   \;=\;
   (-1)^{x_1+x_2} \, (2 \delta_{\sigma(x), \sigma(y)} - 1)
\ee
for diagonal next-nearest-neighbor sites $x,y$.\footnote{
   {\sc Proof:}  Next-nearest-neighbor sites $x,y$ always satisfy
   $h(y) - h(x) = 0$ or $\pm 2$:  the former case occurs when
   $\sigma(x) = \sigma(y)$, and the latter when $\sigma(x) \neq \sigma(y)$.
   It follows that
   $$
      e^{\pm i (\pi/2) [h(y)-h(x)]}  \;=\;  2 \delta_{\sigma(x), \sigma(y)} - 1
      \;.
   $$
   Now multiply this by \reff{h_pi}.
}
It follows that $G=\pi$ corresponds to the staggered polarization.

\bigskip

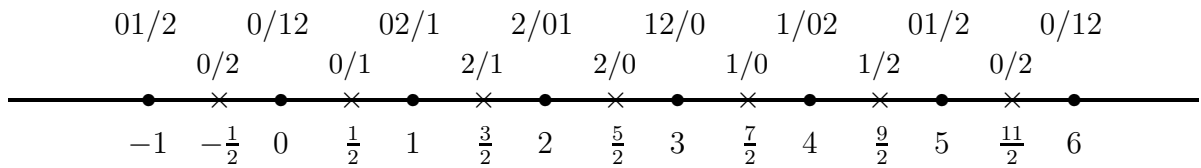
\begin{figure}
\centering
\begin{picture}(450,300)
   \thicklines
   \put(0,200){\line(1,0){450}}
   \put(50,197){$\bullet$}
   \put(75,197){$\times$}
   \put(100,197){$\bullet$}
   \put(125,197){$\times$}
   \put(150,197){$\bullet$}
   \put(175,197){$\times$}
   \put(200,197){$\bullet$}
   \put(225,197){$\times$}
   \put(250,197){$\bullet$}
   \put(275,197){$\times$}
   \put(300,197){$\bullet$}
   \put(325,197){$\times$}
   \put(350,197){$\bullet$}
   \put(375,197){$\times$}
   \put(400,197){$\bullet$}
   \put(45,180){$-1$}
   \put(72,180){$\small -{1 \over 2}$}
   \put(100,180){0}
   \put(127,180){$\small {1 \over 2}$}
   \put(150,180){1}
   \put(177,180){$\small {3 \over 2}$}
   \put(200,180){2}
   \put(227,180){$\small {5 \over 2}$}
   \put(250,180){3}
   \put(277,180){$\small {7 \over 2}$}
   \put(300,180){4}
   \put(327,180){$\small {9 \over 2}$}
   \put(350,180){5}
   \put(374,180){$\small {11 \over 2}$}
   \put(400,180){6}
   \put(40,225){01/2}
   \put(71,210){\small 0/2}
   \put(90,225){0/12}
   \put(121,210){\small 0/1}
   \put(140,225){02/1}
   \put(171,210){\small 2/1}
   \put(190,225){2/01}
   \put(221,210){\small 2/0}
   \put(240,225){12/0}
   \put(271,210){\small 1/0}
   \put(290,225){1/02}
   \put(321,210){\small 1/2}
   \put(340,225){01/2}
   \put(371,210){\small 0/2}
   \put(390,225){0/12}
\end{picture}
\caption{
   The average height $\widetilde{h}(x)$ on a plaquette
   is uniquely determined modulo 6 by the spin content of that plaquette:
   0/12 means, for example,
   that the two spins belonging to the even sublattice are both equal to 0,
   while the two spins belonging to the odd sublattice are 1 and 2.
}
\label{figure_graph2}
\end{figure}

{\bf Remark.}
It is also of interest to define almost-local operators
living on plaquettes.
Let $x$ be a lattice site,
and let
$\Box(x) \equiv \{ (x_1, x_2), \, (x_1 +1, x_2), \,
                   (x_1 +1, x_2 +1), \, (x_1, x_2 +1)  \}$
be the plaquette whose lower-left corner is $x$.
We then define the average height $\widetilde{h}(x)$ over that plaquette as 
\be 
\label{def_htilde} 
\widetilde{h}(x) \;\equiv\; {1 \over 4} \sum_{y \in \Box(x)} h(y) \; .  
\ee
It is easy to see that $\widetilde{h}(x)$ takes values in
${\bf Z} \cup ({\bf Z} + {1 \over 2})$:
namely, it takes an integer (resp.\ half-integer)
value if there are three (resp.\ two)
distinct spin values $\sigma(y)$ on the plaquette $\Box(x)$.
Indeed, the value of $\widetilde{h}(x)$ is uniquely determined mod 6
by the spin content of the plaquette:
see Figure~\ref{figure_graph2}.
Finally, for two adjacent plaquettes $\Box(x)$ and $\Box(x')$,
we have
\be 
\Delta \widetilde{h} \;\equiv\; \widetilde{h}(x') - \widetilde{h}(x)  \;=\;
\cases{ 0,\pm {1\over 2},\pm 1 & if $\widetilde{h}(x) \in {\bf Z}$ \cr   
        \noalign{\vskip 2mm}
        0,\pm {1\over 2}       & if $\widetilde{h}(x) \in {\bf Z} + {1\over 2}$
      } 
\ee
The upshot of this construction is that,
because $\widetilde{h}(x)$ takes half-integral as well as integral values,
vertex operators
$\exp[i G \widetilde{h}(x)]$ and $\exp[i G' \widetilde{h}(x)]$
are equivalent only if $G = G'$ mod $4\pi$,
rather than mod $2\pi$ as before;
so we can define operators up to $|G| = 2\pi$ rather than only $|G| = \pi$.
(However, as will be seen below, all these ``extra'' operators are
irrelevant.)
We have
\begin{eqnarray}
  e^{\pm i (\pi/3) \widetilde{h}(x)}  & = &
     {\sum\limits_{y \in \Box(x)} M_{stagg}^{(1)}(y) \mp i M_{stagg}^{(2)}(y)
      \over
      \; \left| \sum\limits_{y \in \Box(x)}
                M_{stagg}^{(1)}(y) \mp i M_{stagg}^{(2)}(y)
      \right|\;
     }
  \;=\;
     {\sum\limits_{y \in \Box(x)} M_{stagg}^{(1)}(y) \mp i M_{stagg}^{(2)}(y)
      \over
      3 \,+\, (2\sqrt{3}-3) \delta_{\sigma(x),\sigma(x'')}
                            \delta_{\sigma(x'),\sigma(x''')}
     }
       \label{htilde_piover3}   \nonumber \\ \\[4mm]
  e^{\pm i (2\pi/3) \widetilde{h}(x)} & = &
     {\sum\limits_{y \in \Box(x)} M_{u}^{(1)}(y) \pm i M_{u}^{(2)}(y)
      \over
      \;\left| \sum\limits_{y \in \Box(x)} M_{u}^{(1)}(y) \pm i M_{u}^{(2)}(y)
      \right|\;
     }
  \;=\;
     {\sum\limits_{y \in \Box(x)} M_{u}^{(1)}(y) \pm i M_{u}^{(2)}(y)
      \over
      1 \,+\, \delta_{\sigma(x),\sigma(x'')}
                            \delta_{\sigma(x'),\sigma(x''')}
     }
       \label{htilde_2piover3}   \\[4mm]
  e^{\pm i \pi \widetilde{h}(x)}      & = &
%
%
      (-1)^{x_1 + x_2}
     \left[ \delta_{\sigma(x),\sigma(x'')} - \delta_{\sigma(x'),\sigma(x''')} 
            \,\pm\, i 
            \delta_{\sigma(x),\sigma(x'')} \delta_{\sigma(x'),\sigma(x''')}
            \Delta(\sigma(x')-\sigma(x)) \right]  
       \label{htilde_pi}         \nonumber \\  \\[4mm]
  e^{\pm i (4\pi/3) \widetilde{h}(x)}      & = &  
  e^{\pm i (2\pi) \widetilde{h}(x)} \, e^{\mp i (2\pi/3) \widetilde{h}(x)} 
  \;=\; 
  {\sum\limits_{y \in \Box(x)} M_{u}^{(1)}(y) \mp i M_{u}^{(2)}(y)
      \over
      1 \,-\, 3 \delta_{\sigma(x),\sigma(x'')}
                            \delta_{\sigma(x'),\sigma(x''')}
     }
       \label{htilde_4piover3}         \\[4mm]
  e^{\pm i (5\pi/3) \widetilde{h}(x)}      & = &  
  e^{\pm i (2\pi) \widetilde{h}(x)} \, e^{\mp i (2\pi/3) \widetilde{h}(x)}
  \;=\;
     {\sum\limits_{y \in \Box(x)} M_{stagg}^{(1)}(y) \pm i M_{stagg}^{(2)}(y)
      \over
      3 \,-\, (2\sqrt{3}+3) \delta_{\sigma(x),\sigma(x'')}
                            \delta_{\sigma(x'),\sigma(x''')}
     }
       \label{htilde_5piover3}         \\[4mm]
  e^{\pm i (2\pi) \widetilde{h}(x)}      & = &
       1 - 2\delta_{\sigma(x),\sigma(x'')} \delta_{\sigma(x'),\sigma(x''')}
       \label{htilde_2pi}
\end{eqnarray}
where we have labelled the sites around the plaquette $\Box(x)$
as $x, x', x'', x'''$ in cyclic order,
and in \reff{htilde_pi} we have used the shorthand
$\Delta(n) = \pm 1$ according as $n = \pm 1$ mod~3.

\bigskip

We can now read off the predictions for critical exponents.
The staggered magnetization corresponds to $G = \pi/3 = a_{{\cal R}^\circ}$
(this is the smallest nonzero vector in ${\cal R}^\circ$),
hence $\eta_{{\bf M}_{stagg}} = \pi/(18 K)$.
On the other hand,
den Nijs {\em et al.}\/ \cite{Nijs_82}
and Park and Widom \cite{Park_89}
obtained the exact value $\eta_{{\bf M}_{stagg}} = 1/3$
by means of a mapping to the 6-vertex model.
It follows that the height model corresponding to the
3-state square-lattice Potts antiferromagnet at zero temperature
has stiffness $K=\pi/6$.
(In particular, we have $K < K_r = \pi/2$,
 so the height model lies in its rough phase.)
By the usual scaling law we obtain the susceptibility exponent
$(\gamma/\nu)_{stagg} = 2 - \eta_{{\bf M}_{stagg}} = 5/3$.
This value has been numerically verified by several authors
\cite{Wang_89,Wang_90,Ferreira_Sokal_95,Ferreira_Sokal_prep}.

The uniform magnetization
corresponds to $G=2 \pi/3 = 2a_{{\cal R}^\circ}$.
(In this model the ideal states have a nonzero net magnetization,
 which, however, is the same for A/BC and BC/A;
 the uniform magnetization is thus
 periodic on the ideal-state lattice with period 3.)
It follows that $\eta_{{\bf M}_u} = 4 \eta_{{\bf M}_{stagg}} = 4/3$
and $(\gamma/\nu)_u = 2 - \eta_{{\bf M}_{u}} = 2/3$ \cite{Nijs_82}.
It is interesting that the {\em uniform}\/ magnetization
is predicted to have a divergent susceptibility
in this {\em anti}\/ferromagnetic model.
We are not aware of any numerical test of this
prediction in the literature.

The staggered polarization corresponds to
$G=\pi = 3a_{{\cal R}^\circ}$.
We have $\eta_{{\bf P}_{stagg}}= 9 \eta_{{\bf M}_{stagg}} = 3$
and hence $(\gamma/\nu)_{{\bf P}_{stagg}} = 2- \eta_{{\bf P}_{stagg}} = -1$
\cite{Burton_Henley_97}.
This means that the ``susceptibility'' for this operator does not diverge,
but tends to a finite value with a power-law correction $\sim L^{-1}$
(where $L$ is the linear lattice size).
This prediction has not, to our knowledge,
been checked numerically in the literature.

These are the {\em only}\/ relevant vertex operators in the model.
Indeed, a vertex operator $\exp[i G\cdot \bar{h}(x)]$
is relevant if and only if $\eta = G^2/(2 \pi K) < 4$;
or, writing $G = n a_{{\cal R}^\circ}$ with $n$ integer,
we need $|n| < \sqrt{8\pi K} / a_{{\cal R}^\circ}$.
The values $K=\pi/6$ and $a_{{\cal R}^\circ} = \pi/3$
then imply that we must have $|G| < 2 \pi /\sqrt{3}$,
or $|n| < \sqrt{12}$.

The equivalence lattice has lattice spacing $a_{\cal E}=1$,
so that the wavevector corresponding to the locking potential is
$G=2 \pi = 6 a_{{\cal R}^\circ}$
and hence $\eta_{V_{\rm lock}} = 36 \eta_{{\bf M}_{stagg}} = 12 > 4$.
So, $V_{\rm lock}$ is a (strongly) irrelevant operator.

\bigskip

{\bf Remark.}
The foregoing predictions contain, at first glance, a serious paradox.
The correlation functions of the {\em microscopic}\/ staggered and
uniform magnetizations,
\begin{eqnarray}
   G_{stagg}(x)  & = &
      \langle {\bf M}_{stagg}(0) \cdot {\bf M}_{stagg}(x) \rangle     \\[1mm]
   G_{u}(x)  & = &
      \langle {\bf M}_{u}(0) \cdot {\bf M}_{u}(x) \rangle
\end{eqnarray}
obviously satisfy
\be
   G_{stagg}(x)  \;=\;  (-1)^{x_1 + x_2} \, G_{u}(x)   \;.
\ee
How, then, can $G_{stagg}(x)$ decay at large $|x|$ like $|x|^{-1/3}$
while $G_{u}(x)$ decays like $|x|^{-4/3}$?
The answer, presumably, is that the correlation functions contain
{\em both}\/ terms \cite{Nijs_82}:
\begin{eqnarray}
   G_{stagg}(x)  & \sim &
      \hphantom{(-1)^{x_1 + x_2}} |x|^{-1/3}  \,+\,
      (-1)^{x_1 + x_2} |x|^{-4/3}  \,+\,  \ldots            \\[1mm]
   G_{u}(x)  & \sim &
      (-1)^{x_1 + x_2} |x|^{-1/3}  \,+\,
      \hphantom{(-1)^{x_1 + x_2}} |x|^{-4/3}  \,+\,  \ldots
\end{eqnarray}
It is only when one passes to {\em coarse-grained}\/ correlation functions,
by smearing over several nearby lattice sites,
that the oscillatory terms are replaced by much-more-rapidly decaying
remnants, leaving\footnote{
   In order to get the maximum additional decay
   (namely, two powers of $|x|$),
   it is necessary to smear over an $m \times n$ block
   with $m$ and $n$ both {\em even}\/.
}
\begin{eqnarray}
   \bar{G}_{stagg}(x)  & \sim &
      |x|^{-1/3}  \,+\,
      |x|^{-10/3}  \,+\,  \ldots            \\[1mm]
   \bar{G}_{u}(x)  & \sim &
      |x|^{-7/3}  \,+\,
      |x|^{-4/3}  \,+\,  \ldots
\end{eqnarray}
A similar cancellation of oscillatory terms occurs, of course,
when one looks at the susceptibilities.

\section{Numerical Simulations} \label{sec_numerical_simulations}

In order to test all these predictions, we have carried out
a Monte Carlo simulation of the 3-state square-lattice Potts antiferromagnet
at zero temperature,
on periodic $L \times L$ lattices with $L$ ranging from 4 to 1024.
We made our simulation using the Wang-Swendsen-Koteck\'y (WSK)
cluster algorithm \cite{Wang_89,Wang_90},
which is ergodic at $T=0$ on any bipartite graph,
and in particular on a periodic square lattice
whenever the linear lattice size $L$ is {\em even}\/
\cite{Burton_Henley_97,Ferreira_Sokal_prep}.\footnote{
   By contrast, the WSK algorithm for $q=3$ is known to be {\em non}\/ergodic
   on periodic $3m \times 3n$ square lattices
   whenever $m$ and $n$ are relatively prime \cite{Lubin-Sokal}.
   Other cases are open questions.
}

For each lattice size, we made $10^6$ measurements
after discarding $10^5$ iterations for equilibration.
For $L\leq 512$ we performed a single long run starting from
the ordered state 0/1.
For $L=1024$ we made two independent runs with different initial conditions,
one starting in the ordered state 0/1
and the other starting in the ideal state 0/12
(each individual run was of total length $6 \times 10^5$,
 with the first $10^5$ iterations discarded);
there was no noticeable disagreement between the two sets of results.
In units of the longest autocorrelation time
$\tau_{{\rm int},{\cal P}^2_{stagg}}$ (see below),
our run length corresponds to
$\approx 1.3 \times 10^5 \tau_{\rm int}$ measurements,
and our discard interval corresponds to
$\approx 1.3 \times 10^4 \tau_{\rm int}$ iterations.
This run length is sufficient to get a high-precision determination of
both static and dynamic observables:  we obtain errors of order
$\ltapprox 0.2\%$ for the static observables and of order $\ltapprox 2\%$ for
the dynamic ones.\footnote{
   Our discard interval might seem to be much larger than necessary:
   $10^2 \tau_{\rm int}$ would usually be more than enough.
   However, there is always the danger that the longest autocorrelation time
   in the system is much larger than the longest autocorrelation time
   that one has {\em measured}\/, because one has failed to measure
   an observable having sufficiently strong overlap with the slowest mode.
   (Here is a minor example of this effect:
   the authors of Refs.~\cite{Ferreira_Sokal_95,Ferreira_Sokal_prep}
   reported $\tau_{\rm int} \ltapprox 5$ because they failed to consider
   our slowest observable ${\cal P}_{stagg}$,
   which has autocorrelation time
   $\tau_{{\rm int},{\cal P}^2_{stagg}} \approx 8$.)
   As an undoubtedly overly conservative precaution against the possible
   (but unlikely) existence of such a (vastly) slower mode,
   we decided to discard approximately 10\% of the entire run.
   This discard amounts to reducing the accuracy on our final estimates
   by a mere 5\%.

   Note also that while we have here performed our simulations
   only at zero temperature ($\beta=\infty$),
   the authors of Refs.~\cite{Ferreira_Sokal_95,Ferreira_Sokal_prep}
   employed a closely-spaced set of temperatures ranging from
   very high temperature ($\beta=2.0$, $\xi \approx 5$)
   to very low temperature ($\beta=6.0$, $\xi \approx 20000$)
   and found the autocorrelation times
   of ${\cal M}^2_{stagg}$ and the energy
   to be uniformly small.
   This constitutes further evidence against the existence
   of an undetected extremely slow mode.
}

Our program was written in {\sc Fortran}.
The runs for $L \leq 512$ were carried out on a Pentium 166 machine:
each WSK iteration took approximately $5.7 \ L^2$ $\mu$sec.
The runs for $L=1024$ were carried out on an IBM RS-6000/370 workstation,
taking $8.5\ L^2$ $\mu$sec per iteration.
The total CPU time used in this project was approximately
1 month on the former machine plus 4 months on the latter.

The ``zero-momentum'' observables
\begin{eqnarray}
   {\cal M}_{stagg} & = &   \sum_x {\bf M}_{stagg}(x)    \\[2mm]
   {\cal M}_u       & = &   \sum_x {\bf M}_{u}(x)        \\[2mm]
   {\cal P}_{stagg} & = &   \sum_x {\bf P}_{stagg}(x)
\end{eqnarray}
all have mean zero.
We have therefore measured their squares
\begin{eqnarray}
{\cal M}_{stagg}^2 &=& \left( \sum_x {\bf M}_{stagg}(x) \right)^2   \;=\;
                       {3 \over 2} \sum_a \left|
                  \sum_x (-1)^{x_1 + x_2} \delta_{\sigma_x,a} \right|^2 \\[2mm]
{\cal M}_u^2       &=& \left( \sum_x {\bf M}_u(x) \right)^2   \;=\;
                       {3 \over 2} \sum_a \left|
                  \sum_x  \delta_{\sigma_x,a} \right|^2  - {V^2 \over 2}\\[2mm]
{\cal P}_{stagg}^2 &=& \left( \sum_x {\bf P}_{stagg}(x) \right)^2   \;=\;
                             \left| \sum_{\langle x y \rangle \; {\rm nnn}}
                             (-1)^{x_1+x_2}
                             \delta_{\sigma_x,\sigma_{y}} \right|^2
\end{eqnarray}
as well as the ``smallest-nonzero-momentum'' observable associated
to ${\bf M}_{stagg}(x)$:
\begin{eqnarray}
{\cal F}_{stagg}   &=& {1\over 2} \left\{
                       \left| \sum_x e^{2 \pi i x_1 /L} \, {\bf M}_{stagg}(x)
                              \right|^2 +
                       \left| \sum_x e^{2 \pi i x_2 /L} \, {\bf M}_{stagg}(x)
                              \right|^2 \right\}
             \nonumber \\
                   &=& {3 \over 2} \times {1 \over 2} \sum_a \left\{
                      \left| \sum_x (-1)^{x_1 + x_2} e^{2 \pi i x_1 /L}
                             \delta_{\sigma_x,a} \right|^2 +
                       \left| \sum_x (-1)^{x_1 + x_2} e^{2 \pi i x_2 /L}
                              \delta_{\sigma_x,a} \right|^2 \right\}
                     \;.
             \nonumber \\
\end{eqnarray}
Here $V = L^2$ is the volume of the system, the sum $\sum_a$ is over
the three possible values of the Potts spins, and the sum
$\sum_{\langle x y \rangle\; {\rm nnn}}$ is over all pairs of
diagonal-next-nearest neighbors $x,y$ (each pair taken only once).
The staggered and uniform susceptibilities are given by
\begin{eqnarray}
\chi_{stagg} &=& {1 \over V} \langle {\cal M}_{stagg}^2 \rangle \\[2mm]
\chi_u       &=& {1 \over V} \langle {\cal M}_u^2 \rangle
\end{eqnarray}
and the ``susceptibility'' associated to the observable ${\cal P}_{stagg}$
is
\begin{equation}
\chi_{P_{stagg}}   \;=\;  {1 \over V}
               \langle {\cal P}_{stagg}^2 \rangle  \;.
\end{equation}
Finally, the second-moment correlation length is defined by
\begin{equation}
\label{xi_def}
\xi  \;=\;
       { \displaystyle  [(\chi_{stagg}/F_{stagg}) - 1] ^{1/2}
         \over
         \displaystyle  2 \sin (\pi/L)
       }
    \;,
\end{equation}
where
\begin{equation}
F_{stagg} \;=\; {1 \over V} \langle {\cal F}_{stagg} \rangle
   \;.
\end{equation}
The results of our simulations
for the mean values of all these static observables are displayed in
Table~\ref{Table_static}.

We have also measured the integrated autocorrelation time
associated to each of the basic observables,
using a self-consistent truncation window
of width $6 \tau_{\rm int}$
\cite[Appendix C]{Madras_Sokal_88}.
We find that the largest autocorrelation time
(of the observables we measured)
corresponds to ${\cal P}_{stagg}^2$,
though all of them are roughly of the same order of magnitude
(Table~\ref{Table_dynamic}).
None of these autocorrelation times diverges as $L$ grows;
they tend to a constant.
We have fitted the autocorrelation time for each observable to a constant
(using methods to be described at the beginning of the next section).
Our best fits are:
\begin{eqnarray}
  \tau_{{\rm int},{\cal P}^2_{stagg}}   &=& 7.552 \pm 0.052 \qquad
      (L_{min}=128, \, \chi^2=5.43, \hbox{ 3 DF}) \\
  \tau_{{\rm int},{\cal M}^2_{u}}       &=& 4.921 \pm 0.027 \qquad
      (L_{min}=128, \, \chi^2=1.87, \hbox{ 3 DF}) \\
  \tau_{{\rm int},{\cal M}^2_{stagg}}   &=& 4.528 \pm 0.028 \qquad
      (L_{min}=256, \, \chi^2=0.56, \hbox{ 2 DF}) \\
  \tau_{{\rm int},{\cal F}_{stagg}}     &=& 3.804 \pm 0.015 \qquad
      (L_{min}=32,\phantom{2}  \, \chi^2=4.81, \hbox{ 5 DF})
\end{eqnarray}
We conclude that the WSK algorithm for this
model at $T=0$ has no critical slowing-down
\cite{Ferreira_Sokal_95,Ferreira_Sokal_prep}:
$\tau_{\rm int} \ltapprox 8$ uniformly in $L$.

\section{Data Analysis}   \label{sec_data_analysis}

We perform all fits using the standard weighted least-squares method.
As a precaution against corrections to scaling,
we impose a lower cutoff $L \ge L_{min}$
on the data points admitted in the fit,
and we study systematically the effects of varying $L_{min}$
on both the estimated parameters and the $\chi^2$.
In general, our preferred fit corresponds to the smallest $L_{min}$
for which the goodness of fit is reasonable
(e.g., the confidence level\footnote{
   ``Confidence level'' is the probability that $\chi^2$ would
   exceed the observed value, assuming that the underlying statistical
   model is correct.  An unusually low confidence level
   (e.g., less than 5\%) thus suggests that the underlying statistical model
   is {\em incorrect}\/ --- the most likely cause of which would be
   corrections to scaling.
}
is $\gtapprox$ 10--20\%)
and for which subsequent increases in $L_{min}$ do not cause the
$\chi^2$ to drop vastly more than one unit per degree of freedom.

%
%

\subsection{Staggered Susceptibility}

The theoretically expected behavior of the staggered susceptibility at
criticality (i.e., at zero temperature) is
\begin{equation}
  \chi_{stagg}  \;=\;
   L^{(\gamma/\nu)_{stagg}} \left[ A + B L^{-\Delta} + \ldots \right]
  \label{chi_stagg_ansatz}
\end{equation}
with $(\gamma/\nu)_{stagg}=5/3$;
here $\Delta$ is a correction-to-scaling exponent
and the dots indicate higher-order corrections to scaling.
Based on the numerical results of
Refs.~\cite{Ferreira_Sokal_95,Ferreira_Sokal_prep},
we do not expect large corrections to scaling on this observable.

We tried first to extract the leading term in \reff{chi_stagg_ansatz}
by fitting our data
to a simple power-law Ansatz $\chi_{stagg} = A L^{(\gamma/\nu)_{stagg}}$.
This fit is reasonable already for $L_{min} = 32$
($\chi^2 = 4.34$, 4 DF, level = 36\%),
but our preferred fit is $L_{min}=128$:
\begin{equation}
\left({\gamma \over \nu}\right)_{stagg}  \;=\;   1.66621 \pm 0.00035
\end{equation}
with $\chi^2 = 1.31$ (2 DF, confidence level = 52\%).
This result is only 1.5 standard deviations away
from the expected value 5/3.

We then considered the Ansatz \reff{chi_stagg_ansatz},
imposing the leading exponent $(\gamma/\nu)_{stagg}=5/3$
and trying various values for the
first correction-to-scaling exponent $\Delta$.
We are able to find reasonably good fits already for $L_{min}=4$,
provided we take $\Delta$ in the range
$1.50 \ltapprox \Delta \ltapprox 1.76$.
We therefore performed a three-parameter nonlinear weighted least-squares fit
to simultaneously estimate $A$, $B$ and $\Delta$.
Using $L_{min}= 4$, we obtain
\begin{equation}
  \Delta  \;=\;  1.624 \pm 0.061
 \label{eq4.3}
\end{equation}
with $\chi^2 = 7.30$ (6 DF, level = 29\%).

It is interesting to note that the exponent $5/3$ is included
in the interval \reff{eq4.3}.
If this is the true behavior,
it means that the leading correction to pure power-law behavior
in the staggered susceptibility is merely an additive constant:
\begin{equation}
  \chi_{stagg} \;=\; 0.87696(17) L^{5/3} - 0.2820(30)
 \label{chistagg_const}
\end{equation}
with $\chi^2 = 7.78$ (7 DF, level = 35\%).
Such a correction can be interpreted as a mere lattice artifact,
not necessarily arising from any irrelevant operator of the continuum theory.

%
%

\subsection{Uniform Susceptibility}

The theoretically expected behavior for the uniform susceptibility is
\begin{equation}
  \chi_{u}  \;=\;
     L^{(\gamma/\nu)_{u}} \left[ A + B L^{-\Delta} + \ldots \right]
  \label{chi_u_ansatz}
\end{equation}
with $(\gamma/\nu)_{u} = 2/3$.
The simple power-law Ansatz gives a decent fit only for $L_{min}=256$,
yielding
\begin{equation}
\left({\gamma \over \nu}\right)_u   \;=\;   0.6705 \pm 0.0022
\end{equation}
with $\chi^2=0.35$ (1 DF, level = 55\%). This result is 1.75
standard deviations away from the theoretical prediction.

The large deviations from pure power-law behavior for $L < 256$
can be explained as an effect of corrections to scaling.
Indeed, if we consider the Ansatz \reff{chi_u_ansatz} with
$(\gamma/\nu)_{u} = 2/3$ imposed and with just one
correction-to-scaling term, we can obtain sensible fits even for $L_{min}=4$.
But in this case, in contrast to the preceding one,
the range of acceptable $\Delta$ values is much narrower:
$0.655 \ltapprox \Delta \ltapprox 0.735$.
A three-parameter nonlinear weighted least-squares fit
to $A$, $B$ and $\Delta$, with $L_{min}= 4$, yields
\begin{equation}
   \Delta   \;=\; 0.695 \pm 0.013
  \label{eq4.7}
\end{equation}
with $\chi^2 = 1.96$ (6 DF, level = 92\%).
In this case the value $\Delta=2/3$ is two standard deviations away from the
above estimate, but the absolute discrepancy is small (less than 0.03)
and can plausibly be explained as an effect of
higher-order corrections to scaling.
Indeed, the uniform susceptibility can be fitted well
(with $L_{min} = 4$) as a pure power law plus an additive constant:
\begin{equation}
  \chi_{u}  \;=\;  0.54744(49) L^{2/3} - 0.3486(16)   \;.
 \label{chiu_const}
\end{equation}
with $\chi^2=7.15$ (7 DF, level = 41\%).

The results \reff{eq4.3}/\reff{chistagg_const} and
\reff{eq4.7}/\reff{chiu_const},
taken together, suggest that there are no irrelevant operators
(having the symmetries of the Hamiltonian)
with $\Delta < 5/3$
and that the leading corrections to scaling in both
$\chi_{stagg}$ and $\chi_u$ are lattice artifacts.
This behavior is consistent with the prediction \reff{irrelevant}
that the leading irrelevant operator is $(\nabla \bar{h})^4$,
with $\Delta = 2$.

%
%

\subsection{Staggered Polarization}

The finite-size-scaling behavior of $\chi_{P_{stagg}}$ is expected to be
\begin{equation}
\label{p_stagg_ansatz}
  \chi_{P_{stagg}}  \;=\; \chi_{P_{stagg}}(\infty) + B L^{-\Delta} + \ldots
\end{equation}
with $\Delta = 1$.
We tried first to ignore the correction-to-scaling term and fit the data
to a constant. The fit is not very good: even for $L_{min}=128$ we have
$\chi^2=6.68$ (3 DF, level = 8\%), with the estimate
\begin{equation}
  \chi_{P_{stagg}}(\infty) \;=\; 2.1736 \pm 0.0060  \;,
\end{equation}
and the confidence level gets slightly worse for $L_{min} = 256, 512$.

We next fit to \reff{p_stagg_ansatz} with $\Delta = 1$.
For $L_{min} = 8$ one already gets a fair (though not spectacular) fit:
\begin{equation}
   \chi_{P_{stagg}}(\infty) \;=\; 2.1728 \pm 0.0054
\end{equation}
with $\chi^2 = 9.63$ (6 DF, level = 14\%).
However, the confidence level does not improve significantly for
larger $L_{min}$.

We also tried fits to \reff{p_stagg_ansatz} with various values
$\Delta \neq 1$.
We were able to get reasonable fits for $L_{min}=8$,
if we take $0.50 \ltapprox \Delta \ltapprox 1.05$.
We then tried a three-parameter fit to obtain estimates for
$\chi_{P_{stagg}}(\infty)$, $B$ and $\Delta$. Our preferred fit
corresponds to $L_{min}=8$:
\begin{subeqnarray}
   \chi_{P_{stagg}}(\infty) &=&   2.160 \pm 0.011 \\
   \Delta                   &=&   0.75  \pm 0.12
\end{subeqnarray}
with $\chi^2=6.39$ (5 DF, level = 27\%).
The discrepancy between the above result and the predicted value $\Delta=1$
is only two standard deviations;
it might be due to higher-order corrections.

%
%

\subsection{Correlation Length}

Finally, we consider the scaling behavior of the second-moment
correlation length, which is expected to be of the form
\begin{equation}
\label{xi_ansatz}
\xi  \;=\;   L^p \left[ x^\star  + B L^{-\Delta} + \ldots \right]
\end{equation}
with $p = 1$.

First, we tried to estimate the power $p$ by a simple power-law fit.
This gives a good result for $L_{min}=128$:
\begin{equation}
p  \;=\;  0.99875 \pm 0.00069
\end{equation}
with $\chi^2=0.48$ (2 DF, level = 79\%).
This estimate is only 1.8 standard deviations
away from the expected value $p=1$,
and the very small discrepancy (less than 0.0013)
can be explained as an effect of corrections to scaling.

If we look at Table~\ref{Table_xioverL},
we see that the ratio $\xi/L$ increases from $L=4$ to $L=8$,
decreases monotonically from $L=8$ to $L=64$,
and then oscillates due to statistical noise for $L > 64$.
Thus, if we want to study the $L\to\infty$ limit of this quantity
{\em without}\/ including correction-to-scaling terms,
we expect to get a reasonable fit only for $L_{min}\geq 64$.
Indeed, if we fit our data to a constant $x^\star$,
the first decent fit occurs for $L_{min}=64$, giving
\be
x^\star  \;=\;  0.63546 \pm 0.00030
\ee
with $\chi^2=3.77$ (4 DF, level = 44\%). However, our preferred
fit corresponds to $L_{min}=512$,
\be
x^\star  \;=\;  0.63483 \pm 0.00048
\ee
with $\chi^2=0.0051$ (1 DF, level = 94\%).

On the other hand, if we want to study corrections to scaling,
we must use at least some of the data with $L\leq 64$.
The non-monotonic behavior for $4 \le L \le 64$
indicates that, to obtain a reasonable fit over this whole interval,
we would need at least {\em two}\/ correction-to-scaling terms
with amplitudes of {\em opposite}\/ sign.
An Ansatz with only one correction-to-scaling term could,
at best, fit the data with $L_{min}\geq 8$, and very likely not even that.

Indeed, if we fit the data to the Ansatz \reff{xi_ansatz} with $p=1$ and
only one correction-to-scaling term $\sim L^{-\Delta}$,
we find that reasonably good fits are obtained for
$0.25 \ltapprox \Delta \ltapprox 0.60$ with $L_{min}=8$.
(For $L_{min}=4$ we were unable to find any good fit, as expected.)
We next tried a three-parameter fit to estimate
$x^\star$, $B$ and $\Delta$. The first reasonably good fit corresponds
again to $L_{min}=8$, and the estimates are
\begin{subeqnarray}
   x^\star   &=&  0.63359 \pm 0.00132 \\
   \Delta    &=&  0.42    \pm 0.16
     \slabel{Lmin_8_delta}
\end{subeqnarray}
with $\chi^2=9.35$ (5 DF, level = 10\%).
However, a better fit is obtained with $L_{min}=16$, giving
\begin{subeqnarray}
\slabel{result_xstar_1}
   x^\star   &=&  0.63479 \pm 0.00066   \\
   \Delta    &=&  0.84     \pm 0.32
     \slabel{Lmin_16_delta}
\end{subeqnarray}
with $\chi^2=5.95$ (4 DF, level = 20\%).
For $L_{min}\geq 32$, we do not get any sensible result
($\Delta$ and $B$ become very large, along with their error bars);
this is due to the fact that most of these data correspond to
the regime $L \ge 64$ where the corrections to scaling
are submerged under the statistical noise.
Let us remark that
the value of $x^\star$ given in \reff{result_xstar_1} is only
$1.6$ standard deviations away
from the one estimated
by Ferreira and Sokal \cite{Ferreira_Sokal_prep}
using extrapolation techniques at nonzero temperature:
\begin{equation}
  x^\star_{FS}  \;\approx\;  0.633888 \; .
\end{equation}

If we want to fit {\em all}\/ the data (i.e.\ take $L_{min}=4$),
we should introduce at least two correction-to-scaling terms.
{}From the definition \reff{xi_def}, we expect two
types of corrections to scaling for the correlation length: one of order
$L^{-5/3}$ coming from the numerator [cf.\ \reff{chistagg_const}],
and another of order $L^{-2}$ coming from the
subleading terms in the sine.
Furthermore, we might also expect an effective  constant-term ``correction''
of order $L^{-1}$, analogously to what happened for the two susceptibilities.
Thus, our next Ansatz would be
\begin{equation}
\label{xi_ansatz_3}
  {\xi \over L}  \;=\;   x^\star  + B L^{-1} + C L^{-5/3}   \;.
\end{equation}
If the coefficients $B$ and $C$ have different signs,
the contribution of these two terms could be mimicked
(in the range of monotonicity, $L_{min} \ge 8$)
by a single correction term with an exponent $\Delta_{eff} < 1$,
where $\Delta_{eff}$ increases towards 1 as $L_{min} \to\infty$.
Indeed, this scenario is in good agreement with our results
\reff{Lmin_8_delta}/\reff{Lmin_16_delta}.
We therefore tried a three-parameter fit directly
to the Ansatz \reff{xi_ansatz_3}.
Our preferred fit corresponds to $L_{min}=4$:
\begin{subeqnarray}
x^\star  &=& \phantom{-}0.63457  \pm 0.00033 \\
B        &=& \phantom{-}0.182    \pm 0.014 \\
C        &=&          -0.521     \pm 0.035
\end{subeqnarray}
with $\chi^2 = 7.01$ (6 DF, level = 32\%).
This result certainly does not {\em prove}\/
that the Ansatz \reff{xi_ansatz_3} is correct,
since many other pairs of correction-to-scaling exponents
could give an equally good fit;
but it does display a satisfying agreement.

%
%
\section*{Acknowledgments}

We wish to thank Chris Henley for valuable correspondence and
for making available some of his unpublished notes.

J.S.\ gratefully acknowledges the hospitality of the Department of Physics
at New York University, where this work was finished.
The authors' research was supported in part by
CICyT grants PB95-0797 and AEN97-1680 (J.S.)
and by U.S.\ National Science Foundation grant PHY-9520978 (A.D.S.\ and J.S.).

\newpage
\renewcommand{\baselinestretch}{1}
\large\normalsize
%
%
%
%

\clearpage

\newpage

\def\tt{\phantom{1}}
\def\ttt{\phantom{-}}
%
%
\begin{table}[htb]
\centering
\begin{tabular}{|l|c|c|c|l|}
\hline\hline\\[-0.5cm]
\multicolumn{1}{c|}{Operator}&
\multicolumn{1}{c|}{$G$}&
\multicolumn{1}{c|}{$\eta$}  &
\multicolumn{1}{c|}{$\gamma/\nu = 2 - \eta$} &
\multicolumn{1}{c|}{Numerical Result} \\
\hline\hline
${\bf M}_{stagg}$ & $\pm \pi/3$  & $1/3$ & $5/3$ & $\ttt1.66621   \pm 0.00035$
   \\
${\bf M}_u$       & $\pm 2\pi/3$ & $4/3$ & $2/3$ & $\ttt0.6705\tt \pm 0.0022\tt$
   \\
${\bf P}_{stagg}$ & $\pm \pi$    & 3     & $-1$  & $   -0.75\tt\tt\tt \pm 0.12$
   \\
\hline\hline
\end{tabular}
\caption{Critical operators for the 3-state antiferromagnetic Potts model
         on the square lattice at zero temperature.
         The last column indicates the results from our Monte Carlo simulation
         (Section~\protect\ref{sec_data_analysis}).
        }
\label{Table_operators}

\end{table}

%
%
\begin{table}[htb]
\hspace*{-1cm}
\begin{tabular}{|r|r|r|r|r|}
\hline\hline\\[-0.5cm]
\multicolumn{1}{c|}{$L$}&
\multicolumn{1}{c|}{$\chi_{stagg}$}&
\multicolumn{1}{c|}{$\chi_u$}&
\multicolumn{1}{c|}{$\chi_{P_{stagg}}$}&
\multicolumn{1}{c|}{$\xi$}\\
\hline\hline
   4  & $    8.5576 \pm \tt 0.0024$ & $    1.0443 \pm    0.0010$ & $ 2.4486 \pm
0.0076$ & $   2.5127 \pm    0.0023$ \\
   8  & $   27.7671 \pm \tt 0.0120$ & $    1.8614 \pm    0.0028$ & $ 2.3888 \pm
0.0124$ & $   5.1306 \pm    0.0055$ \\
  16  & $   88.8206 \pm \tt 0.0441$ & $    3.1537 \pm    0.0058$ & $ 2.2882 \pm
0.0128$ & $  10.2497 \pm    0.0110$ \\
  32  & $  282.8085 \pm \tt 0.1477$ & $    5.2052 \pm    0.0105$ & $ 2.2464 \pm
0.0126$ & $  20.4422 \pm    0.0219$ \\
  64  & $  897.3520 \pm \tt 0.4776$ & $    8.4362 \pm    0.0179$ & $ 2.2110 \pm
0.0124$ & $  40.6662 \pm    0.0433$ \\
 128  & $ 2851.2642 \pm \tt 1.5484$ & $   13.5899 \pm    0.0296$ & $ 2.1826 \pm
0.0120$ & $  81.4388 \pm    0.0880$ \\
 256  & $ 9056.7475 \pm \tt 4.8888$ & $   21.8141 \pm    0.0479$ & $ 2.1852 \pm
0.0122$ & $ 162.8235 \pm    0.1745$ \\
 512  & $28731.6260 \pm   15.5273$ & $   34.7753 \pm    0.0767$ & $ 2.1480 \pm
0.0116$ & $ 325.0478 \pm    0.3449$ \\
1024  & $91167.3235 \pm   49.2042$ & $   55.2604 \pm    0.1240$ & $ 2.1806 \pm
0.0120$ & $ 650.0264 \pm    0.6924$ \\
\hline\hline
\end{tabular}
\caption{Mean values of the static observables for the 3-state square-lattice
          Potts
          antiferromagnet at zero temperature.}
\label{Table_static}
\end{table}

%
%
\begin{table}[hbt]
\centering
\begin{tabular}{|r|r|r|r|r|}
\hline\hline\\[-0.5cm]
\multicolumn{1}{c|}{$L$}&
\multicolumn{1}{c|}{$\tau_{{\rm int},{\cal M}_{stagg}^2}$}&
\multicolumn{1}{c|}{$\tau_{{\rm int},{\cal M}_{u}^2}$}&
\multicolumn{1}{c|}{$\tau_{{\rm int},{\cal P}_{stagg}^2}$}&
\multicolumn{1}{c|}{$\tau_{{\rm int},{\cal F}_{stagg}}$}\\
\hline\hline
   4  & $    2.487 \pm    0.020$ & $    1.525 \pm    0.010$ & $    5.038 \pm    0.057$ & $    3.330 \pm    0.030$\\
   8  & $    4.073 \pm    0.041$ & $    3.240 \pm    0.029$ & $    7.456 \pm    0.101$ & $    4.032 \pm    0.041$\\
  16  & $    4.417 \pm    0.046$ & $    4.154 \pm    0.042$ & $    7.845 \pm    0.109$ & $    3.850 \pm    0.038$\\
  32  & $    4.488 \pm    0.047$ & $    4.628 \pm    0.049$ & $    7.765 \pm    0.107$ & $    3.826 \pm    0.037$\\
  64  & $    4.483 \pm    0.047$ & $    4.803 \pm    0.052$ & $    7.863 \pm    0.110$ & $    3.803 \pm    0.037$\\
 128  & $    4.587 \pm    0.049$ & $    4.903 \pm    0.054$ & $    7.528 \pm    0.103$ & $    3.855 \pm    0.038$\\
 256  & $    4.558 \pm    0.049$ & $    4.908 \pm    0.054$ & $    7.686 \pm    0.106$ & $    3.816 \pm    0.037$\\
 512  & $    4.515 \pm    0.048$ & $    4.889 \pm    0.054$ & $    7.377 \pm    0.100$ & $    3.752 \pm    0.036$\\
1024  & $    4.512 \pm    0.048$ & $    4.986 \pm    0.056$ & $    7.638 \pm    0.105$ & $    3.776 \pm    0.037$\\
\hline\hline
\end{tabular}
\caption{Mean values of the dynamic observables for the 3-state square-lattice
          Potts
          antiferromagnet at zero temperature.}
\label{Table_dynamic}
\end{table}

%
%
\begin{table}[htb]
\centering
\begin{tabular}{|r|r|}
\hline\hline\\[-0.5cm]
\multicolumn{1}{c|}{$L$}&
\multicolumn{1}{c|}{$\xi/L$}\\
\hline\hline
   4  & $  0.62818 \pm  0.00057$ \\
   8  & $  0.64133 \pm  0.00069$ \\
  16  & $  0.64061 \pm  0.00069$ \\
  32  & $  0.63882 \pm  0.00068$ \\
  64  & $  0.63541 \pm  0.00068$ \\
 128  & $  0.63624 \pm  0.00069$ \\
 256  & $  0.63603 \pm  0.00068$ \\
 512  & $  0.63486 \pm  0.00067$ \\
1024  & $  0.63479 \pm  0.00068$ \\
\hline\hline
\end{tabular}
\caption{Values of the ratio $\xi/L$ for the 3-state
          square-lattice Potts antiferromagnet at zero temperature.}
\label{Table_xioverL}
\end{table}


\newpage


%
%

%
%


\end{document}